\documentclass[11pt]{article}
\usepackage[letterpaper,margin=1in]{geometry}
\usepackage{mathptmx}
\usepackage{amsmath,amssymb,bm}
\usepackage{graphicx}
\usepackage[protrusion=true,expansion=false]{microtype}
\usepackage[font=small,labelfont=bf,labelsep=period,margin=0.5in]{caption}
\usepackage{subcaption}
\usepackage{titlesec}
\usepackage{authblk}
\usepackage{xcolor}
\usepackage[colorlinks=true,linkcolor={blue!50!black},citecolor={blue!50!black},urlcolor={blue!50!black}]{hyperref}

\titleformat{\section}{\large\bfseries}{\thesection.}{0.6em}{}
\titleformat{\subsection}{\normalsize\bfseries}{\thesubsection.}{0.6em}{}
\titlespacing*{\section}{0pt}{2.2ex plus .5ex}{1ex}
\titlespacing*{\subsection}{0pt}{1.8ex plus .4ex}{0.8ex}
\newcommand{\Elm}{\mathrm{Elm}}
\newcommand{\Ekin}{E_{\mathrm{kin}}}
\newcommand{\ceff}{c^{\mathrm{eff}}}
\newcommand{\fd}{\hat{f}^{\dagger}}
\newcommand{\f}{\hat{f}}
\newcommand{\vin}{\mathbf{in}}
\newcommand{\vout}{\mathbf{out}}
\newcommand{\vElm}{\mathbf{Elm}}
\newcommand{\dEdR}[1]{\tfrac{\mathrm{d}E}{\mathrm{d}R_{#1}}}
\newcommand{\Bt}{\overleftrightarrow{B}}

\title{\Large\bfseries Implementation of LDA\,+\,Gutzwiller with Newton's method}
\author[1]{Jian Zhang}
\author[2,3]{Ming-Feng Tian}
\author[4]{Guang-Xi Jin}
\author[1]{Yuan-Feng Xu}
\author[1,*]{Xi Dai}
\affil[1]{Beijing National Laboratory for Condensed Matter Physics, Institute of Physics, Chinese Academy of Sciences, Beijing 100190, China}
\affil[2]{LCP, Institute of Applied Physics and Computational Mathematics, Beijing 100088, China}
\affil[3]{CAEP Software Center for High Performance Numerical Simulation, Beijing 100088, China}
\affil[4]{Institute of Nuclear Physics and Chemistry, China Academy of Engineering Physics, Mianyang 621900, China}
\date{}

\begin{document}
\maketitle

\vspace{-5ex}
\begin{center}\footnotesize
$^{*}$Corresponding author. E-mail: daix@aphy.iphy.ac.cn\\[0.6ex]
Authors' version of the article published in \emph{Chinese Physics B} \textbf{26}, 017103 (2017),
doi:\href{https://doi.org/10.1088/1674-1056/26/1/017103}{10.1088/1674-1056/26/1/017103}.
\end{center}

\begin{abstract}
In order to calculate the electronic structure of correlated materials, we propose
implementation of the LDA+Gutzwiller method with Newton's method. The self-consistence
process, efficiency and convergence of calculation are improved dramatically by using
Newton's method with golden section search and other improvement approaches. We compare
the calculated results by applying the previous linear mix method and Newton's method. We
have applied our code to study the electronic structure of several typical strong correlated
materials, including SrVO$_3$, LaCoO$_3$, and La$_2$O$_3$Fe$_2$Se$_2$. Our results fit quite
well with the previous studies.
\end{abstract}

\begin{center}\small
\begin{minipage}{0.85\linewidth}
\end{minipage}
\end{center}

\section{Introduction}
The density function theory (DFT) with local density approximation (LDA) have been
successfully applied to calculate the band structures and related properties of many
materials. However, the LDA fails to capture the strongly correlated nature of strongly
correlated materials because it is based on a single particle picture. In order to solve this
problem, several computational methods have been proposed, including LDA\,+\,$U$,
LDA\,+\,DMFT, and LDA\,+\,Gutzwiller.

LDA\,+\,Gutzwiller method has already been proposed and successfully utilized to investigate
many important strongly correlated materials, such as ferromagnetic metals, iron
superconductors, and topological Kondo insulators.\cite{deng2008,deng2009,zhao2012,wang2015}
Compared to LDA\,+\,DMFT, LDA\,+\,Gutzwiller is much faster and cheaper. However, the
convergence of the self-consistent procedure is the bottle neck for this method, which
prevents it from being applied to much bigger systems.

In this paper, we present an important improvement of the Gutzwiller method by combining it
with Newton's method (a numerical method to determine a function's root) and golden section
search (a technique for finding the extremum (minimum or maximum) of a unimodal function). Our
hybrid method dramatically improves the convergence, and can make the LDA\,+\,Gutzwiller
calculations much faster and stable.

The rest of this paper is organized as follows. In Section~\ref{sec:deriv}, the theory behind the
Gutzwiller method is introduced. In Section~\ref{sec:impl}, implementation details are discussed,
and the linear mix method and Newton's method are compared. Section~\ref{sec:example} details
experimental results of our proposed method. Finally, Section~\ref{sec:concl} lists a short
summary of our findings.

\section{LDA+Gutzwiller derivation}\label{sec:deriv}
In this section, we present the basic derivation of the Gutzwiller method. More detailed
derivations and explanations can be found in our previous paper.\cite{deng2008,deng2009}

\subsection{Hubbard model}
In this section, we derive the basic equations of the Gutzwiller approximation using the
multi-orbital Hubbard model. The Hubbard model can be divided into two parts, one is the
single particle ($H_{\mathrm{lda}}=H_{\mathrm{kin}}+H_{\mathrm{loc}}$) part, which can be
obtained from LDA calculations, the other is the interaction part ($H_{\mathrm{int}}$), which
describes the electron--electron interaction by many-particle wave functions. Expanded
expression is given as follows:
\begin{align}
H &= H_{\mathrm{kin}}+H_{\mathrm{loc}}+H_{\mathrm{int}} \nonumber\\
  &= \sum_{i\neq j,\alpha\beta} t_{i,j}^{\alpha\beta}\,\fd_{i,\alpha}\f_{j,\beta}
   + \sum_{i,\alpha} t_{i,i}^{\alpha\alpha}\,\fd_{i,\alpha}\f_{j,\alpha}
   + \sum_{i} H_{i,\mathrm{int}}, \label{eq:H}\\
H_{\mathrm{int}} &= U_1\sum_{m}\hat{n}_{m\uparrow}\hat{n}_{m\downarrow}
   + U_2\sum_{m<m'}\hat{n}_{m\uparrow}\hat{n}_{m'\downarrow}
   + (U_2-J)\sum_{m<m',\sigma}\hat{n}_{m\sigma}\hat{n}_{m'\sigma} \nonumber\\
 &\quad - J\sum_{m<m'}\left(f^{\dagger}_{m\uparrow}f_{m\downarrow}f^{\dagger}_{m'\uparrow}f_{m'\downarrow}
   + f^{\dagger}_{m\uparrow}f^{\dagger}_{m\downarrow}f_{m'\uparrow}f_{m'\downarrow}+\mathrm{h.c.}\right). \label{eq:Hint}
\end{align}
In Eq.~(\ref{eq:H}), Greek letter subscripts denote a combined spin--orbital index, and the
operator $\fd_{i,\alpha}$ ($\f_{i,\alpha}$) creates (annihilates) an electron on orbital
$\alpha$ at site $i$. The first term of Eq.~(\ref{eq:H}) describes electrons hopping between
sites $i$ and $j$, while the second term describes electrons hoping on one site between
different orbitals. The last term describes the electron--electron interaction. In
Eq.~(\ref{eq:Hint}), $m$ denotes strictly orbital index, while $\sigma$ denotes spin index. The
former three terms are density--density interaction terms: the first term describes
interactions between two electrons in the same orbital but with different spins; the second
term describes interactions between two electrons on different orbitals with different spins
($U_2<U_1$); and the third term describes interactions between two electrons with the same
spin but on different orbitals ($U_2-J<U_2<U_1$). The third term's factor is the smallest, thus
electrons with the same spin tend to have lower system energy. This reflects Hund's rules. The
fourth term describes interactions of spin flipping and pair hopping.

In LDA\,+\,Gutzwiller method, two types of interaction terms can be considered. One is called
density--density Gutzwiller interaction (ddgw), in which only the first three terms in
Eq.~(\ref{eq:Hint}) are included. There are two parameters affecting the energy band, the
quasi-particle weight $R$ and the onsite energy revision Elm. $R$ is between 0 and 1, and the
band is compressed according to $R$. Elm affects the shift of the correlated bands. In this
type, quasi-particle weight $R$ and on-site energy Elm are vectors $R_\alpha$ and
$\Elm_\alpha$. $\alpha$ is the local orbital index. The other type, rotational invariable
Gutzwiller interaction (rtgw), includes all interacting terms. In rtgw, quasi-particle weight
and on-site energy are matrix $\bm{R}_{\alpha,\beta}$ and $\Elm_{\alpha,\beta}$. $\alpha$ and
$\beta$ are local orbital index. In the present paper, we will focus on the density--density
type just for simplicity. The generalization to the rotational invariant type is straight
forward. The result of the rotational invariant type is shown in Section~\ref{sec:lacoo3}.

\subsection{Gutzwiller variational wave function}
The Gutzwiller wave function (GWF) $|G\rangle$\cite{bunemann1998} is a many-particle wave
function. We act the Gutzwiller projection operator $\hat{P}$ upon the Hartree uncorrelated
wave function (HWF) $|0\rangle$
\begin{equation}
|G\rangle=\hat{P}|0\rangle.
\end{equation}
This operator $P$ is composed by projector $P_i$
\begin{equation}
\hat{P}=\prod_i\hat{P}_i,
\end{equation}
where $i$ is the site index, and $P_i$ projects the wave function to many-particle configurations
\begin{equation}
\hat{P}_i=\sum_I c_I|I\rangle\langle I|.
\end{equation}
Here, $I$ denotes the Fock states, the parameters $c_I$ are variational parameters, which
adjust the weight of each configuration in the wave function. The GWF is obtained by setting
the proper parameters $c_I$.

\subsection{Variational problem constraints}
There are two variational problem constraints. First, the GWF $|G\rangle$ must be normalized
as follows:
\begin{equation}
\langle G|G\rangle=\prod_i\langle 0|\hat{P}_i\hat{P}_i|0\rangle=1.
\end{equation}
The other constraint is\cite{bunemann1998}
\begin{equation}\label{eq:constraint}
\langle 0|\hat{P}_i^{\dagger}\hat{P}_i\fd_{i,\alpha}\f_{i,\alpha}|0\rangle
=\langle 0|\fd_{i,\alpha}\f_{i,\alpha}|0\rangle,
\end{equation}
and we denote
\begin{align}
n^0_\alpha &= \langle 0|\fd_{i,\alpha}\f_{i,\alpha}|0\rangle,\\
N_\alpha &= \langle 0|\hat{P}_i^{\dagger}\hat{P}_i\fd_{i,\alpha}\f_{i,\alpha}|0\rangle.
\end{align}

\subsection{Lagrange equation}
The Lagrange function is $F\left(c_I,\ceff_{nl},\lambda_\varphi,\lambda_{lk},\lambda_\alpha\right)$,
in which $c_I$ is the Lagrange variation parameter to determine the GWF. $\ceff_{nl}$ is the
Lagrange variation parameter to determine effective one-particle wave function $|l\rangle$.
$|n\rangle$ is the eigen wave function from LDA calculation, we use $|n\rangle$ as single
particle basis. $\lambda_{lk}$ and $\lambda_\varphi$ are the Lagrange multipliers of the
normalization constraint for the effective one-particle wave function and the GWF,
respectively. $\lambda_\alpha$ is the Lagrange multiplier of the particle number operator
constraint, and its physical meaning is each orbit's chemical potential.
\begin{align}
F\left(c_I,\ceff_{nl},\lambda_\varphi,\lambda_{lk},\lambda_\alpha\right)
&= \sum_{i,j,\alpha,\beta} t_{i,j}^{\alpha,\beta}\langle G|\fd_{i,\alpha}\f_{j,\beta}|G\rangle
 + \langle G|H_{\mathrm{int}}|G\rangle
 - \sum_{lk}\lambda_{lk}\left(\langle\psi_{lk}|\psi_{lk}\rangle-1\right)\nonumber\\
&\quad - \sum_{\alpha}\lambda_\alpha\left[N_\alpha-n^0_\alpha\right]
 - \lambda_\varphi\left(\langle\varphi|\varphi\rangle-1\right).
\end{align}
In this non-linear equation, both the effective single particle wave function $\ceff_{nl}$ and
the many-particle wave function $c_I$ are unknown, and they are related to each other. To solve
this equation, we designed a two-step self-consistent loop. First, all parameters related to
the many-particle wave function ($R_\alpha$, $\Elm_\alpha$) are set as constants, and we can find
a proper effective single particle wave function $\ceff_{nl}$. Second, all parameters related to
the single particle wave function ($n_\alpha$, $\Ekin/R_\alpha$) are set as constants, and we can
constitute hamiltonian on the non-interacting many-particle wave function and find the proper
variational parameters $c_I$. The equation for the effective single-particle part is as follows:
\begin{align}
\frac{\partial F}{\partial c^{\mathrm{eff}*}_{nl}}
&= \sum_{i,j,\alpha,\beta} t_{i,j}^{\alpha,\beta}
 \frac{\partial\langle G|\fd_{i,\alpha}\f_{j,\beta}|G\rangle}{\partial c^{\mathrm{eff}*}_{nl}}
 + \frac{\partial\sum_\alpha\lambda_\alpha\left[n^0_\alpha-N_\alpha\right]}{\partial c^{\mathrm{eff}*}_{nl}}
 - \lambda_{lk}\ceff_{nl} \nonumber\\
&= 0. \label{eq:single}
\end{align}
The equation for the many-particle part is
\begin{align}
\frac{\partial F}{\partial c_I}
&= \sum_{i,j,\alpha,\beta} t_{i,j}^{\alpha,\beta}
 \frac{\partial\langle G|\fd_{i,\alpha}\f_{j,\beta}|G\rangle}{\partial c_I}
 + \frac{\partial\langle G|H_{\mathrm{int}}|G\rangle}{\partial c_I}\nonumber\\
&\quad - \frac{\partial\left(\sum_\alpha\lambda_\alpha\left[N_\alpha-n^0_\alpha\right]\right)}{\partial c_I}
 - \lambda_\varphi\sum_I 2c_I\langle 0|I\rangle\langle I|0\rangle \nonumber\\
&= 0. \label{eq:many}
\end{align}

\section{Implementation and optimization}\label{sec:impl}
In this section, we focus on implementation details and optimization techniques applied to the
Gutzwiller method. We introduce the interface of LDA\,+\,Gutzwiller method. Then outline the two
self-consistent loops (outer loop and inner loop), and compare the linear mixed method and
Newton's method. Lastly, we discuss the golden section search method used in the inner loop.

\subsection{Interface of LDA+Gutzwiller}
The Gutzwiller method requires two input files. One is the eigenvalue of each $k$ point
$E_{nk}$, which results from LDA calculation. The other is the overlap matrix
$\bm{S}_{\alpha,n}(k)$, $S_{\alpha,n}(k)=\langle\phi_{\alpha k}|\psi_{nk}\rangle$.
$|\psi_{nk}\rangle$ is the eigenfunction from LDA calculation. $|\phi_{\alpha k}\rangle$ is the
local orbital. In our method, strong correlation confined to each site is required, therefore
local orbitals between neighbor sites should not be overlapped. The local orbital can be a
Wannier function\cite{anisimov2005,lechermann2006} of the LDA bloch wave, or atomic local
orbitals.

Tight binding Harmiltonian can be built on local orbital basis. Since some tight binding models
are provided in real space, the Hamiltonian is $H(r)$, and it is better to transform to
reciprocal space $H(k)$ by Fourier transformation, to facilitate the effective single particle
calculation. Details on building local orbital and projection can be found in previous
paper.\cite{zhao2012}

\subsection{Two-step loop of Gutzwiller method}
\begin{figure}[htbp]
\centering
\includegraphics[width=0.5\linewidth]{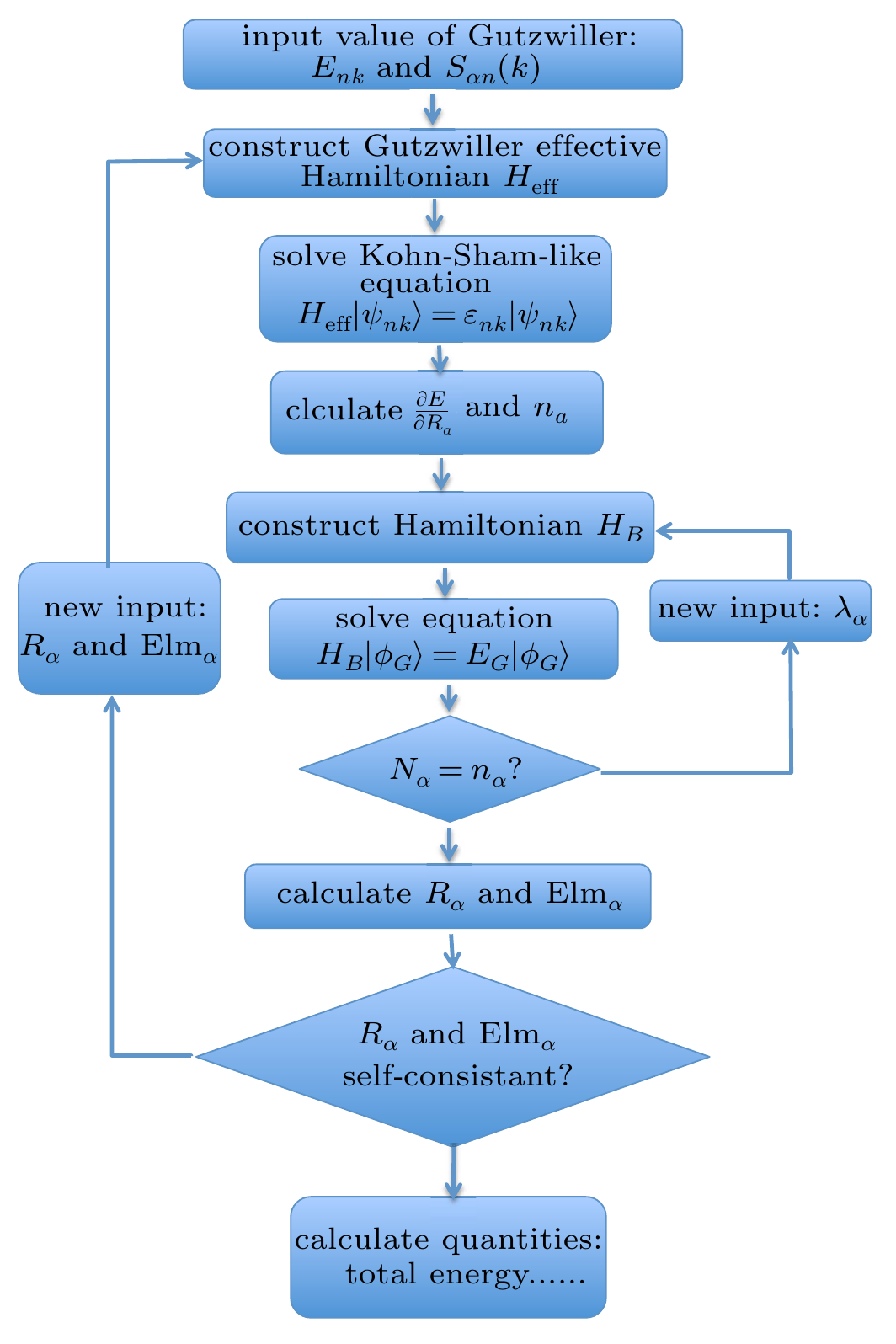}
\caption{Flow chart of the two-step self consistent loop.}\label{fig:flow}
\end{figure}

In the sigle-particle part (Section~\ref{sec:eff}), with given input $R^{\mathrm{input}}_\alpha$
and $\Elm^{\mathrm{input}}_\alpha$, we can find proper $\ceff_{nl}$. Then we can calculate
parameters related with $\ceff_{nl}$, such as occupation number $n_\alpha$ and derivative
$\frac{\partial\Ekin}{\partial R_\alpha}$. In the many-particle part (Section~\ref{sec:mp}), with
occupation number $n_\alpha$ and derivative $\frac{\partial\Ekin}{\partial R_\alpha}$ from
single-particle part, we can find proper $c_I$. The inner self-consistent loop is only related to
many-particle part (Eq.~(\ref{eq:many})). In the inner loop, with given occupation number
$n_\alpha$ and derivative $\frac{\partial\Ekin}{\partial R_\alpha}$, we need to find proper
chemical potential $\lambda_\alpha$ of each orbit to meet the constraint condition
(Eq.~(\ref{eq:constraint})). Then we use occupation number $n_\alpha$, derivative
$\frac{\partial\Ekin}{\partial R_\alpha}$ and chemical potential $\lambda_\alpha$ (from inner
self-consistent loop) to find proper $c_I$. Then we can calculate the output
$R^{\mathrm{output}}_\alpha$ and $\Elm^{\mathrm{output}}_\alpha$. The two-step loop is shown in
Fig.~\ref{fig:flow}.

\subsection{Effective single-particle equation}\label{sec:eff}
From Eq.~(\ref{eq:single}), we get an effective single particle Hamiltonian
\begin{align}
H_{\mathrm{eff}}(k) &= \sum_{\alpha\beta} R^{\dagger}_\alpha t_k^{\alpha,\beta}R_\beta
 |k\alpha\rangle\langle k\beta|
 + \sum_\alpha\left(1-\sum_\alpha R^2_\alpha t_k^{\alpha,\alpha}\right)|k\alpha\rangle\langle k\alpha|
 + \sum_\alpha\Elm_\alpha|\alpha k\rangle\langle\alpha k|,\\
\Elm_\alpha &= \frac{\partial\Ekin}{\partial R_\alpha}\sum_{II'}
 \frac{c_Ic_{I'}S^\alpha_{II'}}{\sqrt{n_\alpha\left(1-n_\alpha\right)}}
 \frac{n_\alpha-0.5}{n_\alpha\left(1-n_\alpha\right)}+\lambda_\alpha,\\
R_\alpha &= \sum_{II'}\frac{S^\beta_{II'}c_Ic_{I'}}{\sqrt{n^0_\beta\left(1-n^0_\beta\right)}}.
\end{align}
$R_\alpha$ is quasi-particle weight, $\Elm_\alpha$ is the chemical potential of local orbital
$\alpha$. Both are functions of $c_I$. $n_\alpha$ is the occupation number of local orbital
($n_\alpha=\langle 0|\fd_\alpha\f_\alpha|0\rangle$), $\frac{\mathrm{d}E}{\mathrm{d}R_\alpha}$ is the
kinetic energy, and $f_{nk}$ is the Fermi dispersion of the effective single particle wave
function. $S^\alpha_{II'}$ is either one or zero, which denotes eliminating an electron on
orbital $\alpha$ in Fock state $I'$, and then mapping to Fock state $I$
($S^\alpha_{II'}=\langle I|\f_\alpha|I'\rangle$).

\subsection{Many-particle equation}\label{sec:mp}
From Eq.~(\ref{eq:many}), we obtain a many-particle Hamiltonian written on the basis of Fock states,
\begin{align}
(H_m)_{II'} &= \sum_\alpha\frac{\partial\Ekin}{\partial R_\alpha}
 \frac{S^\alpha_{II'}+S^{\alpha\dagger}_{II'}}{\sqrt{n^0_\alpha\left(1-n^0_\alpha\right)}}
 + 2H^I_{\mathrm{int}}\delta_{II'} - 2\sum_\alpha\lambda_\alpha N^\alpha_I\delta_{II'},\label{eq:Hm}\\
n_\alpha &= \sum_{l,k} w(k)f_{lk}\langle\psi_{lk}|\alpha k\rangle\langle\alpha k|\psi_{lk}\rangle,\\
\frac{\partial\Ekin}{\partial R_\alpha} &= \sum_k\sum_\beta R_\beta t_k^{\beta\alpha}\chi_{k,\alpha\beta}
 + \sum_\beta R_\beta t_k^{\alpha\beta}\chi_{k,\beta\alpha} - 2R_\alpha t_k^{\alpha\alpha}\chi_{k,\alpha\alpha},
\end{align}
where
\begin{equation}
\chi_{k,\alpha\beta}=\langle 0|\fd_{k,\alpha}\f_{k,\beta}|0\rangle
=\sum_n f_{nk}\langle\psi_{nk}|k\alpha\rangle\langle k\beta|\psi_{nk}\rangle.
\end{equation}
$\frac{\partial\Ekin}{\partial R_\alpha}$ and $n_\alpha$ are two input parameters gotten from
the effective single particle equation. $N^\alpha_I$ denotes the electronic occupation on orbital
$\alpha$ in the Fock state $I$ ($N^\alpha_I=\langle\fd_{i,\alpha}\f_{i,\alpha}\rangle_I$). The
ground state eigen vector of $H_m$ is the Lagrange variational parameter $c_I$. Then we can
calculate $N_\alpha$ ($N_\alpha=\sum_I N^\alpha_I c_I^2$) to see whether the constraint has been
met ($N_\alpha=n_\alpha$). Then we adjust $\lambda_\alpha$ to meet the constraint condition
given by Eq.~(\ref{eq:constraint}).

\subsection{Outer self-consistent loop and Newton's method}
Each self-consistent loop was separated into two parts. In the effective single-particle
equation, input is quasi-particle weight $R_\alpha$ and on-site energy $\Elm_\alpha$, output is
kinetic energy $\frac{\partial\Ekin}{\partial R_\alpha}$ and effective occupation number
$n_\alpha$. Conversely, in the many-particle equation, input is kinetic energy
$\frac{\partial\Ekin}{\partial R_\alpha}$ and occupation number $n_\alpha$, output is
quasi-particle weight $R_\alpha$ and on-site energy $\Elm_\alpha$. The two parts constitute a
complete self-consistent loop. We can view the loop as function of $\bm{g}$, with input
$\bm{R}_{\mathrm{in}}$ and $\vElm_{\mathrm{in}}$, and function results of $\bm{R}_{\mathrm{out}}$
and $\vElm_{\mathrm{out}}$. For convenience, input parameters are called $\vin$, while the output
parameters are called $\vout$
\begin{align}
\bm{g}_R(\bm{R}_{\mathrm{in}},\vElm_{\mathrm{in}}) &= \bm{R}_{\mathrm{out}},\nonumber\\
\bm{g}_{\mathrm{Elm}}(\bm{R}_{\mathrm{in}},\vElm_{\mathrm{in}}) &= \vElm_{\mathrm{out}}.
\end{align}
This equation is simplified as
\begin{equation}
\bm{g}(\vin)=\vout.
\end{equation}
Previously, we use linear mix method, that is,
\begin{equation}
\vin_1=\lambda\bm{g}(\vin_0)+(1-\lambda)\vin_0,
\end{equation}
where $\lambda$ is a real number between 0 and 1.

To make this self-consistent procedure more efficient, we then applied Newton's method. Using
this method, the self-consistent problem (function $g$) is changed to a root finding problem
(function $f$) and defined as
\begin{equation}
\bm{f}(\vin)=\bm{g}(\vin)-\vin.
\end{equation}
The self-consistent point $\bm{g}(\vin_r)=\vin_r$ is the point $\bm{f}(\vin_r)=\bm{0}$. Taking the
derivative at point $\vin_0$, which is near the root point $\vin_r$, we get
\begin{equation}
\bm{f}(\vin_r)-\bm{f}(\vin_0)=\frac{\partial\vout}{\partial\vin}(\vin_r-\vin_0).
\end{equation}
Taking $\bm{f}(\vin_r)=\bm{0}$, $\vin_0$ becomes the next input point for our consistent loop.
This equation becomes
\begin{equation}
-\bm{f}(\vin_0)=\frac{\partial\vout}{\partial\vin}(\vin_r-\vin_0),
\end{equation}
and the next input point is
\begin{equation}
\vin_r=\vin_0-\left(\left.\frac{\partial\vout}{\partial\vin}\right|_{\vin_0}\right)^{-1}\bm{f}(\vin_0),
\end{equation}
$\frac{\partial\vout}{\partial\vin}$ is the Jacobi matrix. To apply Newton's method, we must derive
the Jacobian: $\frac{\partial R_{\mathrm{out}}}{\partial R_{\mathrm{in}}}$,
$\frac{\partial R_{\mathrm{out}}}{\partial\Elm_{\mathrm{in}}}$,
$\frac{\partial\Elm_{\mathrm{out}}}{\partial R_{\mathrm{in}}}$ and
$\frac{\partial\Elm_{\mathrm{out}}}{\partial\Elm_{\mathrm{in}}}$. In order to get this Jacobi
matrix, we break it into two parts: the effective single particle part
$\frac{\partial\bm{n}}{\partial\bm{R}_{\mathrm{in}}}$,
$\frac{\partial\frac{\mathrm{d}\bm{E}}{\mathrm{d}\bm{R}}}{\partial\bm{R}_{\mathrm{in}}}$,
$\frac{\partial\bm{n}}{\partial\vElm_{\mathrm{in}}}$,
$\frac{\partial\frac{\mathrm{d}\bm{E}}{\mathrm{d}\bm{R}}}{\partial\vElm_{\mathrm{in}}}$ (see
Appendix~\ref{app:A}), and the many-particle part
$\frac{\partial\bm{R}_{\mathrm{out}}}{\partial\bm{n}}$,
$\frac{\partial\vElm_{\mathrm{out}}}{\partial\bm{n}}$,
$\frac{\partial\bm{R}_{\mathrm{out}}}{\partial\frac{\mathrm{d}\bm{E}}{\mathrm{d}\bm{R}}}$,
$\frac{\partial\vElm_{\mathrm{out}}}{\partial\frac{\mathrm{d}\bm{E}}{\mathrm{d}\bm{R}}}$ (see
Appendix~\ref{app:B}).

In the many-particle part, there is a intermediate variable $\bm{\lambda}$, and we need
$\frac{\partial\bm{\lambda}}{\partial\bm{n}}$ and
$\frac{\partial\bm{\lambda}}{\partial\frac{\mathrm{d}\bm{E}}{\mathrm{d}\bm{R}}}$ to yield the Jacobi
matrix for the many-particle part. The details for these derivatives are shown in
Appendix~\ref{app:C}.

\subsection{Inner self-consistent loop and Newton's method}
The inner self-consistent loop corresponds to the many-particle part outlined in
Section~\ref{sec:mp}. The purpose is to find the proper $\bm{\lambda}$ in Eq.~(\ref{eq:Hm}). In this
loop, the parameter of many-particle Hamiltonian $H_m$ is the chemical potential $\bm{\lambda}$,
$\bm{n}$, and $\frac{\mathrm{d}\bm{E}}{\mathrm{d}\bm{R}}$ are given by single-particle part
(Eq.~(\ref{eq:single})). With given $\bm{\lambda}$, we can get Hamiltonian $H_m$. Diagonalizing
this Hamiltonian yields the ground state wave function $C_I$ and the corresponding $N_\alpha$,
$N_\alpha=\sum_I N^\alpha_I c_I^2$. We define a function $\bm{g}$, where input is $\bm{\lambda}$
and output is $\bm{N}$, $\bm{g}(\bm{\lambda})=\bm{N}$.

To meet the constraint $\bm{N}=\bm{n}$, we define a new function
\begin{equation}
\bm{f}(\bm{\lambda})=\bm{g}(\bm{\lambda})-\bm{n}.
\end{equation}
The Jacobi matrix for the inner loop is $\frac{\partial\bm{N}}{\partial\bm{\lambda}}$. For
derivation details, please refer to Appendix~\ref{app:C} and Eq.~(\ref{eq:C10}).

\subsection{Golden section search with Newton's method}
We also apply golden section search\cite{press2007} to improve the convergence of our code.
Newton's method sets direction in the parameter space, and then golden section search sets the
step length. Newton's method is very sensitive to the starting point. In some situations, it can
cost more steps or cannot find solution, if Newton's method starts with a bad point. With the
technique of golden section search, in each step, we can find the point nearest to the solution
on the step's Newton's line, which is a better starting point for next step, thus it can improve
the convergence. Results are shown in Section~\ref{sec:la2o3}.

\section{Example}\label{sec:example}
In this section, we show results from SrVO$_3$, LaCoO$_3$ and La$_2$O$_3$Fe$_2$Se$_2$, and compare
the LDA and LDA+G band structures. We will also compare our method to the traditional Gutzwiller
method. Convergence behavior will be shown with respect to convergence steps.

\subsection{SrVO$_3$}
\begin{figure}[htbp]
\centering
\includegraphics[width=0.55\linewidth]{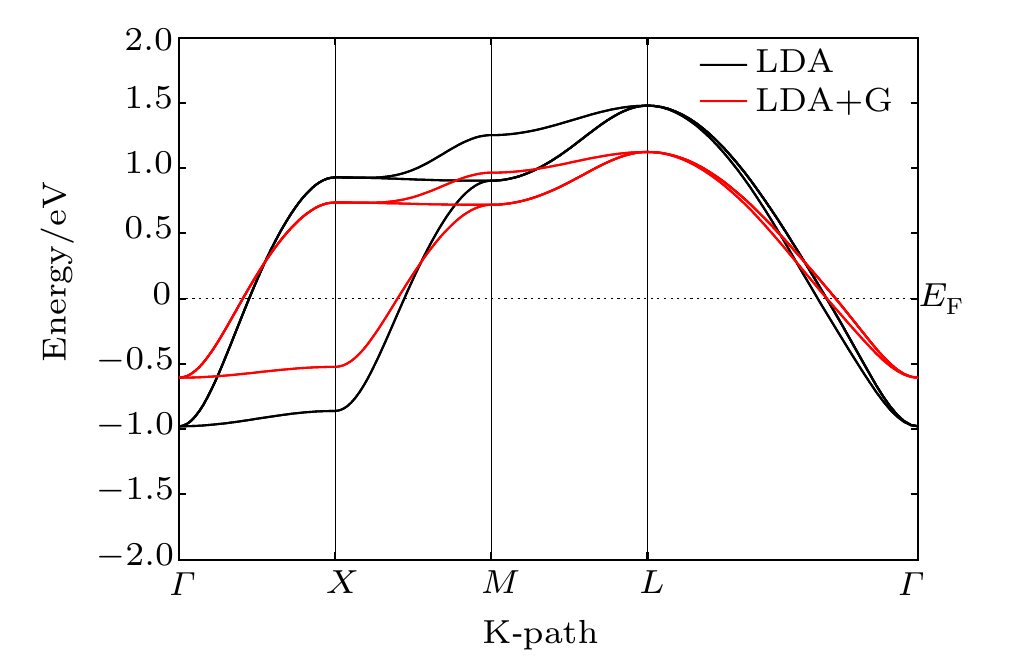}
\caption{The band structure of SrVO$_3$ from the LDA and the LDA\,+\,Gutzwiller result.
$U=5$~eV and $J_z=1.0$~eV.}\label{fig:srvo3band}
\end{figure}

SrVO$_3$ is a 3d orbital correlated metal. Its simple cubic perovskite structure and
non-magnetic electronic state make it an ideal test
material.\cite{lechermann2006,liebsch2003,nekrasov2005} The LDA result is inconsistent with
experiments.\cite{fujimori1992,inoue1995,sekiyama2004} These features can be understood through
our LDA\,+\,Gutzwiller calculations. We choose a 3d$_{t_{2g}}$ Wannier function as the interacting
local orbits with interaction energy $U=5$~eV, $J_z=1.0$~eV. Figure~\ref{fig:srvo3band} shows the
band structure calculated by LDA\,+\,Gutzwiller. The band width using our method is reduced by
30\%. These results are in good agreement with previous
experiments.\cite{fujimori1992,inoue1995,sekiyama2004}

\begin{figure}[htbp]
\centering
\includegraphics[width=0.85\linewidth]{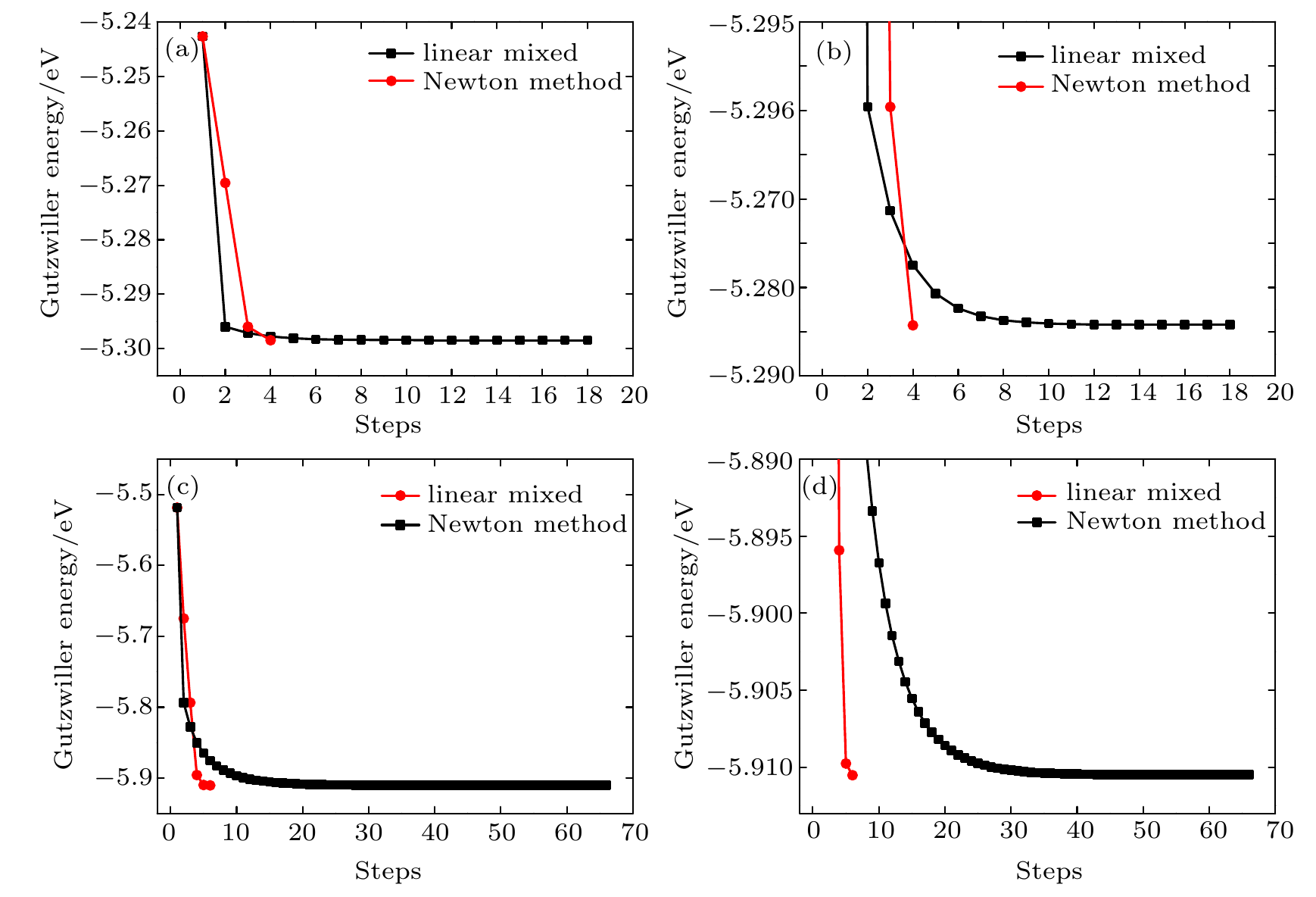}
\caption{The Gutzwiller energy of SrVO$_3$ with respect to steps. Upper case: $U=1$, $J=0$.
Lower case: $U=6$, $J=0$. Figures on the left are zoomed images of the right figures.}
\label{fig:srvo3conv}
\end{figure}

The results of quasi-particle weight and on-site energy by our new application are all well in
comparable with the traditional application. Figure~\ref{fig:srvo3conv} shows convergence
behavior of the two methods in case of SrVO$_3$. Case far away from the transition point
($U=1.0$~eV, $J=0.0$~eV) and near the transition point ($U=6.0$~eV, $J=0.0$~eV) are shown in upper
and lower part respectively. In both cases, applying Newton's method reduces the number of
convergence steps.

\begin{figure}[htbp]
\centering
\includegraphics[width=0.48\linewidth]{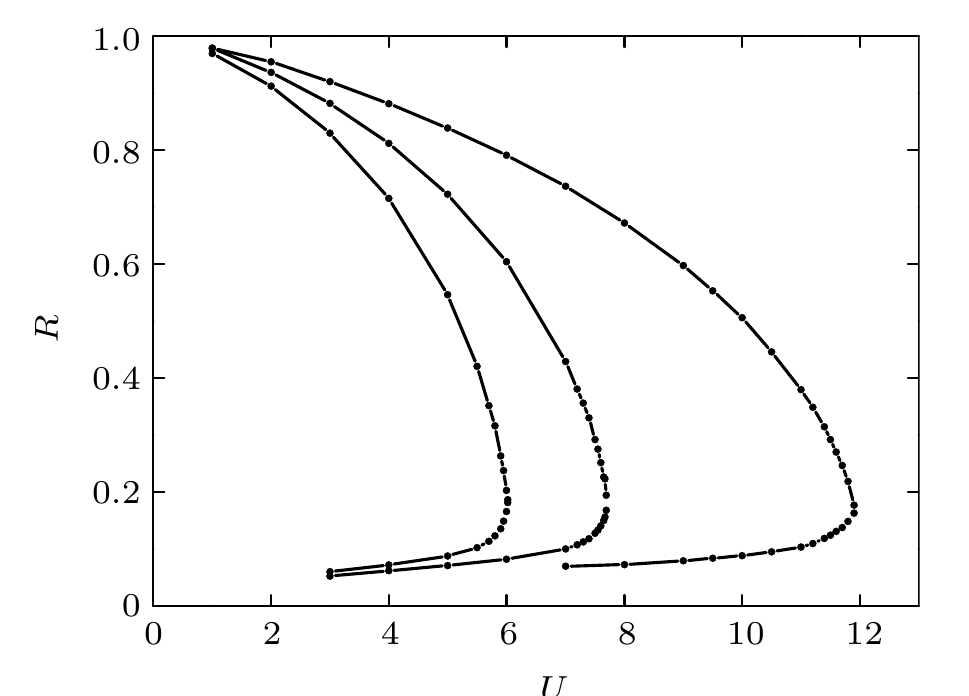}
\caption{The quasi-particle weight $R$\cite{bunemann2003} with respect to $U$. From right to left,
$J/U=0.2$, 0.1, 0.0.}\label{fig:R}
\end{figure}

SrVO$_3$ showed one electron in the 3d orbitals, consistent with previous
findings.\cite{demedici2011} The Brinkman--Rice transition\cite{brinkman1970} occurred at certain
combination of $U$ and $J$. As shown in Fig.~\ref{fig:R}, the Gutzwiller method can converge into
two different results, which is also consistent with previous research.\cite{lanata2012} The
previous linear mix application cannot find self-consistent solution for branches with small
quasi-particle weight. The application of Newton's method can improve the convergence of the
Gutzwiller method.

\subsection{LaCoO$_3$}\label{sec:lacoo3}
\begin{figure}[htbp]
\centering
\includegraphics[width=0.48\linewidth]{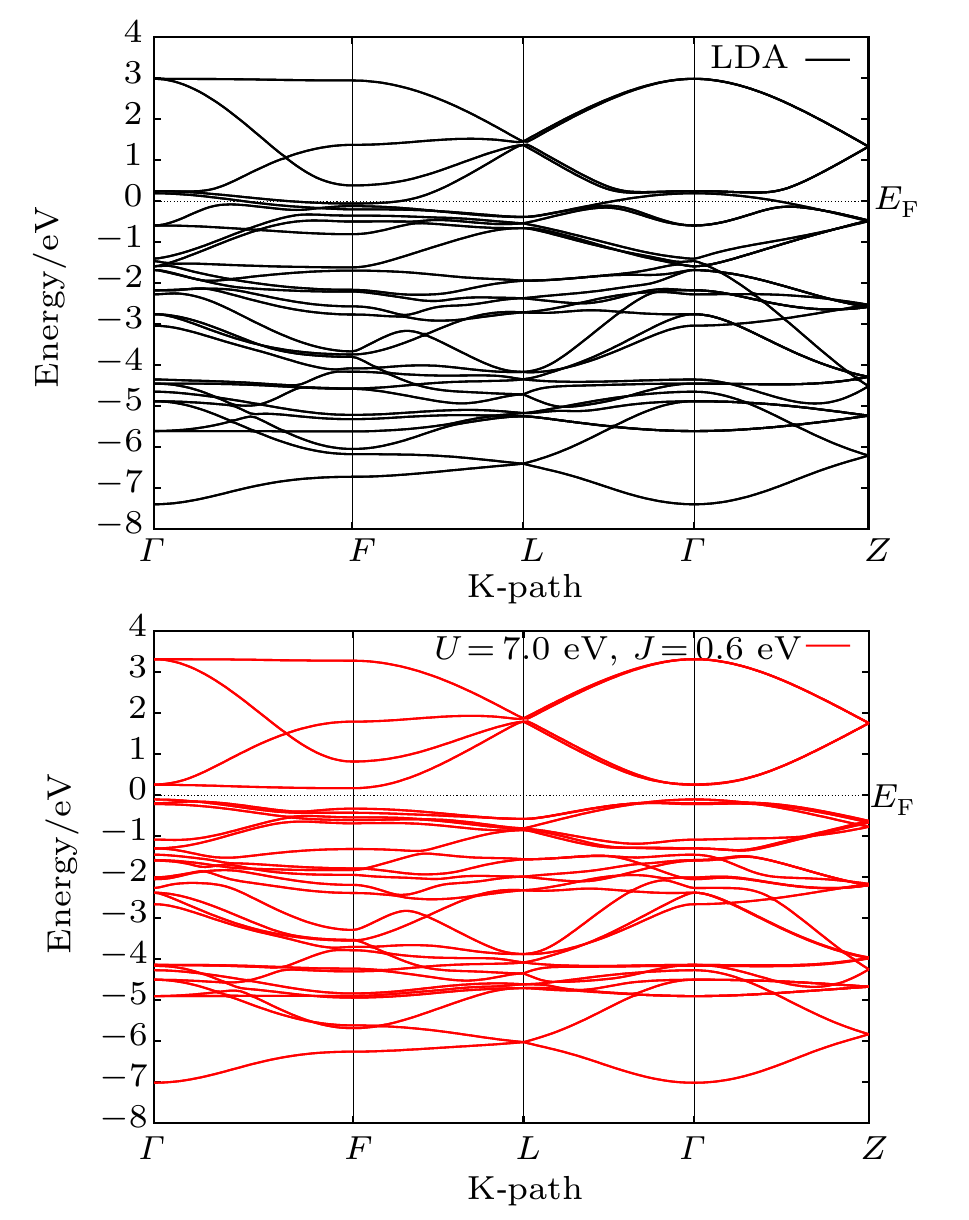}
\caption{LaCoO$_3$ band structure with interaction $U=7.0$~eV, Hund's rule coupling
$J=0.6$~eV.}\label{fig:lacoo3}
\end{figure}

LaCoO$_3$ is a typical correlation material. Since there are two strongly correlated atoms in one
cell, and each atom has 10 local orbitals, convergence of this case is more difficult than
SrVO$_3$. We calculated bulk LaCoO$_3$\cite{wang2015} band structure with LDA and LDA+G (rtgw). We
applied rotational invariant code for this case. The Gutzwiller method adjusted crystal field
splitting between t$_{2g}$ and e$_g$ due to the effect of on-site energy Elm, and renormalized
t$_{2g}$ and e$_g$ bandwidth using quasi-particle weight $R$. LDA and LDA+G bands are plotted in
Fig.~\ref{fig:lacoo3}. The Gutzwiller method shows that LaCoO$_3$ is semiconductor, while LDA
calculation suggests it is metal. The band gap was approximately 0.36~eV, which is in good
agreement with experiments.\cite{chainani1992,abbate1993,arima1993}

\subsection{La$_2$O$_3$Fe$_2$Se$_2$}\label{sec:la2o3}
\begin{figure}[htbp]
\centering
\includegraphics[width=0.6\linewidth]{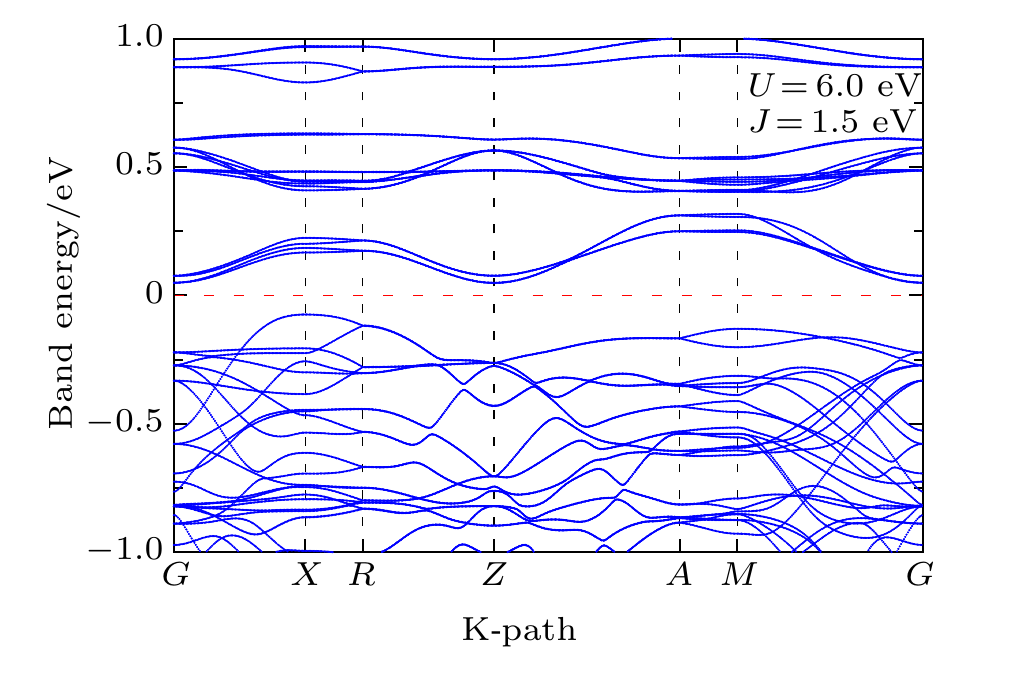}
\caption{La$_2$O$_3$Fe$_2$Se$_2$ band structure with interaction $U=6.0$~eV, Hund's rule coupling
$J=1.5$~eV.}\label{fig:la2o3band}
\end{figure}

La$_2$O$_3$Fe$_2$Se$_2$ is considered a strongly correlated material and is claimed to be a
semiconductor and Mott insulator.\cite{free2010} We consider the magnetic order of
La$_2$O$_3$Fe$_2$Se$_2$. Each cell has 16 strongly correlate atoms, and each atom has 10 local
orbitals. This makes convergence very difficult. The band structure with interaction $U=6.0$~eV
and $J=1.5$~eV is shown in Fig.~\ref{fig:la2o3band}. The band gap was approximately 0.121~eV, which
agrees with past findings.\cite{zhu2010,lei2012,jin2016}

\begin{figure}[htbp]
\centering
\begin{minipage}[t]{0.48\linewidth}
\centering
\includegraphics[width=\linewidth]{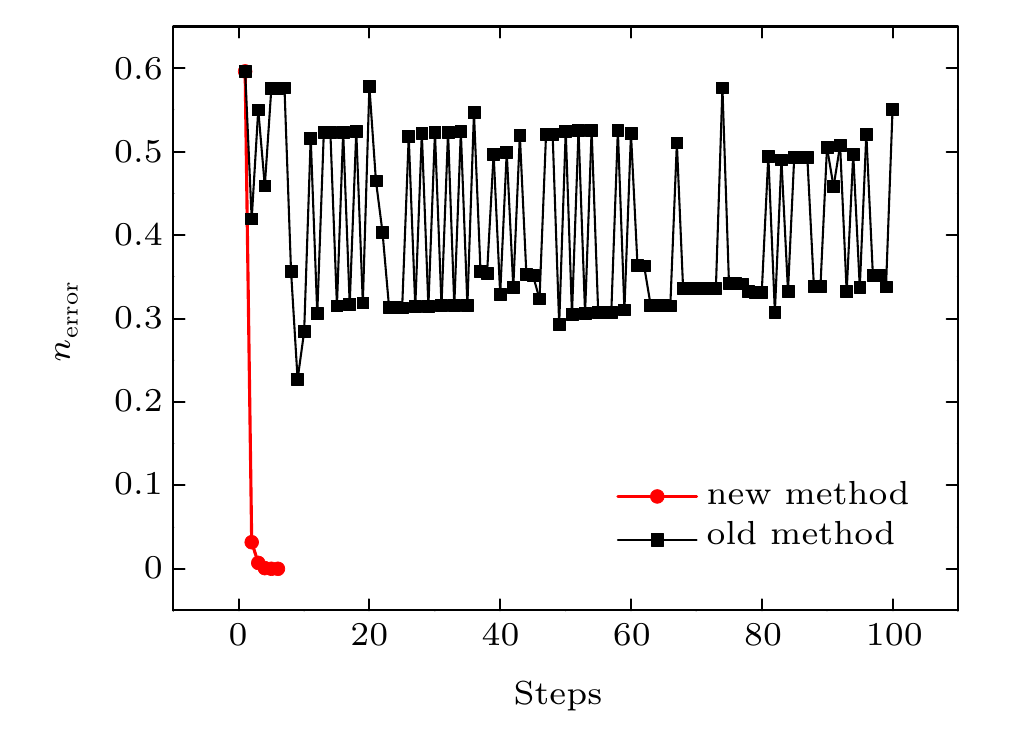}
\captionsetup{margin=0pt}
\caption{The average absolute deviation of the difference between $N_\alpha$ and $n^0_\alpha$
with respect to number of steps.}\label{fig:nerror}
\end{minipage}\hfill
\begin{minipage}[t]{0.48\linewidth}
\centering
\includegraphics[width=\linewidth]{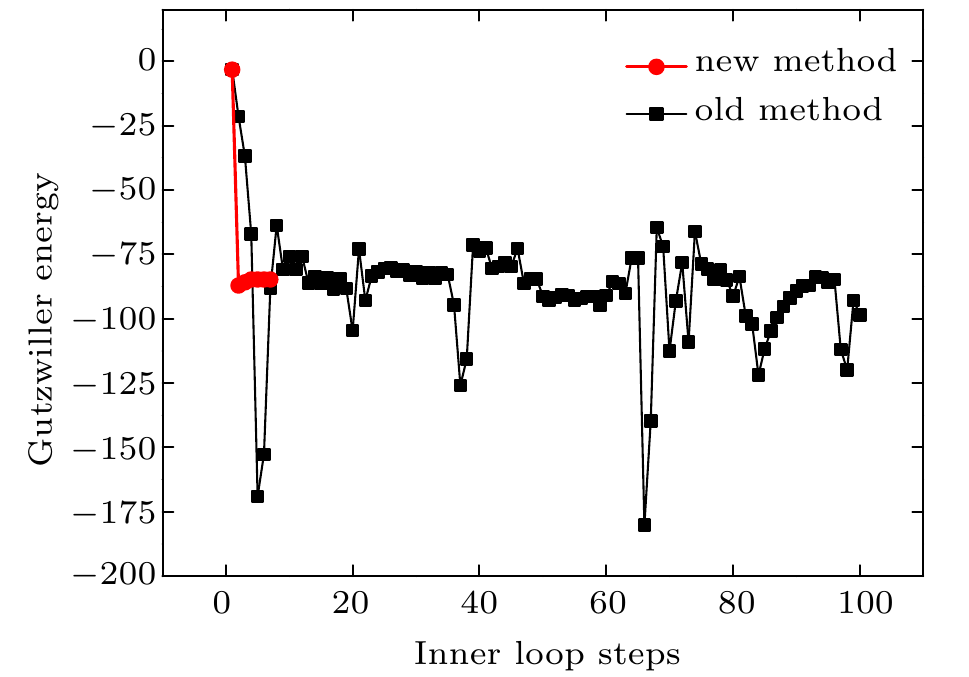}
\captionsetup{margin=0pt}
\caption{The Gutzwiller total energies in the inner self-consistent loop with respect to inner
loop steps, implementation of linear mix and Newton's method respectively.}\label{fig:inner}
\end{minipage}
\end{figure}

Calculations were performed using both linear mix and Newton's methods. The linear mix method
cannot achieve convergence. In the inner loop, the linear mix method cannot locate appropriate
chemical potentials for each orbital $\lambda_\alpha$ defined by Eq.~(\ref{eq:many}) with the
constraint condition given by Eq.~(\ref{eq:constraint}). In contrast, Newton's method locates
appropriate chemical potentials within 6 steps. Figure~\ref{fig:nerror} shows the average absolute
deviation of the difference between $N_\alpha$ and $n^0_\alpha$
$\left(n_{\mathrm{error}}=\sum_\alpha\frac{|N_\alpha-n^0_\alpha|}{N_{\mathrm{orb}}}\right)$ with
respect to number of steps. Figure~\ref{fig:inner} shows the Gutzwiller energy in the inner loop
with respect to the number of steps.

\section{Conclusion}\label{sec:concl}
Our implementation of LDA+G with Newton's method, golden section search, and other revision of
previous Gutzwiller implementation were detailedly described in this paper. First, the Hubbard
model (Section 2.1) and Gutzwiller wave function (Section 2.2) were introduced. Then, we put
forward a two-step self-consistent method to solve the nonlinear Lagrange equation
(Section 2.4), and apply Newton's method to both the outer and inner loops (Sections 3.5 and 3.6
respectively). Finally, we tested our method on SrVO$_3$, LaCoO$_3$, and La$_2$O$_3$Fe$_2$Se$_2$.
Results for all materials agree with both the experimental data and previous calculation.
Conclusively, this proposed method dramatically improves the efficiency and convergence of the
implementation of Gutzwiller method.

\section*{Acknowledgments}
Project supported by the National Natural Science Foundation of China (Grant
No.~2011CBA00108), the National Basic Research Program of China (Grant No.~2013CB921700), and the
Foundation of LCP.

\appendix
\renewcommand{\theequation}{\thesection\arabic{equation}}

\section{Jacobi matrix for single-particle part equation}\label{app:A}
\setcounter{equation}{0}
This is the Jacobi matrix for Newton's method in the effective single-particle equation, in which
input is $\bm{R}$ and $\vElm$, output is $\frac{\mathrm{d}\bm{E}}{\mathrm{d}\bm{R}}$ and $\bm{n}$.
The four elements of the Jacobi matrix are as follows:
\begin{align}
\frac{\partial\dEdR{\alpha}}{\partial R_\beta}
&= \sum_{lk}w_kf_{lk}\sum_{mn}\frac{\partial c^{\mathrm{eff}*}_{lm}}{\partial R_\beta}
 \left(\frac{\partial\Ekin}{\partial R_\alpha}\right)_{mn}\ceff_{nl}
 + w_kf_{lk}\sum_{mn}c^{\mathrm{eff}*}_{lm}\left(\frac{\partial\Ekin}{\partial R_\alpha}\right)_{mn}
 \frac{\partial\ceff_{nl}}{\partial R_\beta}\nonumber\\
&\quad + w_k\frac{\partial f_{lk}}{\partial R_\beta}\sum_{mn}c^{\mathrm{eff}*}_{lm}S^{*}_{m\alpha}S_{\alpha n}\ceff_{nl}
 + w_kf_{lk}\sum_{mn}c^{\mathrm{eff}*}_{lm}\left(\frac{\partial\Ekin}{\partial R_\alpha\partial R_\beta}\right)_{mn}\ceff_{nl}.
\end{align}
Here, $S^{*}_{m\alpha}=\langle m|\alpha\rangle$, $|\alpha\rangle$ is the local orbital, and
$\langle m|$ is the input LDA wave function.
\begin{align}
\frac{\partial\dEdR{\alpha}}{\partial\Elm_\beta}
&= \sum_{lk}w_kf_{lk}\sum_{mn}\frac{\partial c^{\mathrm{eff}*}_{lm}}{\partial\Elm_\beta}
 \left(\frac{\partial\Ekin}{\partial R_\alpha}\right)_{mn}\ceff_{nl}
 + w_kf_{lk}\sum_{mn}c^{\mathrm{eff}*}_{lm}\left(\frac{\partial\Ekin}{\partial R_\alpha}\right)_{mn}
 \frac{\partial\ceff_{nl}}{\partial\Elm_\beta}\nonumber\\
&\quad + w_k\frac{\partial f_{lk}}{\partial\Elm_\beta}\sum_{mn}c^{\mathrm{eff}*}_{lm}S^{*}_{m\alpha}S_{\alpha n}\ceff_{nl},\\
\frac{\partial n_\alpha}{\partial R_\beta}
&= \sum_{lk}w_kf_{lk}\sum_{mn}\frac{\partial c^{\mathrm{eff}*}_{lm}}{\partial R_\beta}S^{*}_{m\alpha}S_{\alpha n}\ceff_{nl}
 + w_kf_{lk}\sum_{mn}c^{\mathrm{eff}*}_{lm}S^{*}_{m\alpha}S_{\alpha n}\frac{\partial\ceff_{nl}}{\partial R_\beta}\nonumber\\
&\quad + w_k\sum_{mn}\frac{\partial f_{lk}}{\partial R_\beta}c^{\mathrm{eff}*}_{lm}S^{*}_{m\alpha}S_{\alpha n}\ceff_{nl},\\
\frac{\partial n_\alpha}{\partial\Elm_\beta}
&= \sum_{lk}w_kf_{lk}\sum_{mn}\frac{\partial c^{\mathrm{eff}*}_{lm}}{\partial\Elm_\beta}S^{*}_{m\alpha}S_{\alpha n}\ceff_{nl}
 + w_kf_{lk}\sum_{mn}c^{\mathrm{eff}*}_{lm}S^{*}_{m\alpha}S_{\alpha n}\frac{\partial\ceff_{nl}}{\partial\Elm_\beta}\nonumber\\
&\quad + w_k\sum_{mn}\frac{\partial f_{lk}}{\partial\Elm_\beta}c^{\mathrm{eff}*}_{lm}S^{*}_{m\alpha}S_{\alpha n}\ceff_{nl}.
\end{align}

\section{Jacobi matrix for many-particle equation}\label{app:B}
\setcounter{equation}{0}
This is the Jacobi matrix for Newton's method in the many-particle equation, in which input is
$\frac{\mathrm{d}\bm{E}}{\mathrm{d}\bm{R}}$, and output is $\bm{R}$, $\vElm$, and $\bm{n}$. The four
elements of the Jacobi matrix are as follows:
\begin{align}
\frac{\partial R_\alpha}{\partial n_\beta}
&= \delta_{\alpha\beta}\sum_{II'}\frac{c_Ic_{I'}S^\alpha_{II'}}{\sqrt{n^0_\alpha\left(1-n^0_\alpha\right)}}
 \frac{n_\alpha-0.5}{n_\alpha\left(1-n_\alpha\right)},\\
\frac{\partial\Elm_\alpha}{\partial n_\beta}
&= \frac{\partial\lambda_\alpha}{\partial n_\beta}
 + \frac{\mathrm{d}E}{\mathrm{d}R_\alpha}\frac{\partial R_\alpha}{\partial n_\beta}
 \frac{n_\alpha-0.5}{n_\alpha\left(1-n_\alpha\right)}f_{nk}|k\alpha\rangle\langle k\alpha|
 + \frac{\mathrm{d}E}{\mathrm{d}R_\alpha}R_\alpha
 \left(\frac{n^2_\alpha-n_\alpha+0.5}{\left(n_\alpha\left(1-n_\alpha\right)\right)^2}\right)
 f_{nk}|k\alpha\rangle\langle k\alpha|,\\
\frac{\partial R_\alpha}{\partial\dEdR{\beta}}
&= \sum_{II'}\left(\frac{\partial c_I}{\partial\dEdR{\beta}}c_{I'}R^\alpha_{II'}
 + c_I\frac{\partial c_{I'}}{\partial\dEdR{\beta}}R^\alpha_{II'}\right),\\
\frac{\partial\Elm_\alpha}{\partial\dEdR{\beta}}
&= \frac{\partial\lambda_\alpha}{\partial\dEdR{\beta}}
 + \left(R_\alpha+\frac{\mathrm{d}E}{\mathrm{d}R_\alpha}\frac{\partial R_\alpha}{\partial\dEdR{\beta}}\right)
 \frac{n_\alpha-0.5}{n_\alpha\left(1-n_\alpha\right)}f_{nk}|k\alpha\rangle\langle k\alpha|.
\end{align}

\section{Details for chemical potential \texorpdfstring{$\bm{\lambda}$}{lambda}}\label{app:C}
\setcounter{equation}{0}
In the many-particle equation, parameter $\bm{\lambda}$ must be adjusted to meet the occupation
number constraint Eq.~(\ref{eq:constraint}). When Newton's method is used for the many-particle
equation, we must know the partial derivatives $\frac{\partial\lambda_\alpha}{\partial n_\beta}$,
$\frac{\partial\lambda_\alpha}{\partial\dEdR{\beta}}$ and
$\frac{\partial N_\alpha}{\partial\lambda_\beta}$.

The occupation number in the many-particle equation is
\begin{equation}
N_\alpha=\sum_I c_I^2N^\alpha_I.
\end{equation}
The differentiation with respect to $c_I$ is
\begin{equation}
\delta N_\alpha=\sum_I 2c_IN^\alpha_I\delta c_I.
\end{equation}
According to the perturbation theory, we can get
\begin{align}
\delta c_I
&= \sum_\alpha\sum_{j'\neq j}
 \frac{\left\langle j'\left|\frac{\partial H_B}{\partial\dEdR{\alpha}}\right|j\right\rangle}{E_j-E_{j'}}
 \langle I|j'\rangle\delta\frac{\mathrm{d}E}{\mathrm{d}R_\alpha}
 + \sum_\alpha\sum_{j'\neq j}
 \frac{\left\langle j'\left|\frac{\partial H_B}{\partial n_\alpha}\right|j\right\rangle}{E_j-E_{j'}}
 \langle I|j'\rangle\delta n_\alpha\nonumber\\
&\quad + \sum_\alpha\sum_{j'\neq j}
 \frac{\left\langle j'\left|\frac{\partial H_B}{\partial\lambda_\alpha}\right|j\right\rangle}{E_j-E_{j'}}
 \langle I|j'\rangle\delta\lambda_\alpha,
\end{align}
define $\left(A_{\dEdR{\alpha}}\right)_I=\sum_{j'\neq j}
\frac{\left\langle j'\left|\frac{\partial H_B}{\partial\dEdR{\alpha}}\right|j\right\rangle}{E_j-E_{j'}}
\langle I|j'\rangle$,
\begin{align}
\delta N_\alpha
&= \sum_\beta\sum_I\left(A_{\dEdR{\beta}}\right)_I 2c_IN^\alpha_I\delta\frac{\mathrm{d}E}{\mathrm{d}R_\beta}
 + \sum_\beta\sum_I\left(A_{n_\beta}\right)_I 2c_IN^\alpha_I\delta n_\beta
 + \sum_\beta\sum_I\left(A_{\lambda_\beta}\right)_I 2c_IN^\alpha_I\delta\lambda_\beta, \label{eq:C4}
\end{align}
define $B^{N^\alpha_I}_{\dEdR{\beta}}=\sum_I\left(A_{\dEdR{\beta}}\right)_I 2c_IN^\alpha_I
\delta\frac{\mathrm{d}E}{\mathrm{d}R_\beta}$,
\begin{align}
\delta N_\alpha &= \sum_\beta\left(B^{N^\alpha_I}_{\dEdR{\beta}}\delta\frac{\mathrm{d}E}{\mathrm{d}R_\beta}
 + B^{N^\alpha_I}_{n_\beta}\delta n_\beta + B^{N^\alpha_I}_{\lambda_\beta}\delta\lambda_\beta\right),\\
\delta\vec{n} &= \delta\vec{N} = \Bt_{\frac{\mathrm{d}E}{\mathrm{d}R}}\,\delta\frac{\mathrm{d}\vec{E}}{\mathrm{d}R}
 + \Bt_n\,\delta\vec{n} + \Bt_\lambda\,\delta\vec{\lambda},\\
\delta\vec{\lambda} &= \left(\Bt_\lambda\right)^{-1}
 \left(\left(1-\Bt_n\right)\delta\vec{n}-\Bt_{\frac{\mathrm{d}E}{\mathrm{d}R}}\,\delta\frac{\mathrm{d}\vec{E}}{\mathrm{d}R}\right),\\
\frac{\partial\lambda_\alpha}{\partial n_\beta} &= \left(\Bt_{\lambda_\alpha}\right)^{-1}_\beta
 - \left(\sum_\gamma\left(\Bt_{\lambda_\alpha}\right)^{-1}_\gamma\Bt_{n_\gamma}\right)_\beta,\\
\frac{\partial\lambda_\alpha}{\partial\dEdR{\beta}} &= -\sum_\gamma\left(\Bt_{\lambda_\alpha}\right)^{-1}_\gamma
 \left(\Bt_{\frac{\mathrm{d}E}{\mathrm{d}R_\gamma}}\right)_\beta.
\end{align}
From Eq.~(\ref{eq:C4}), we determine the inner loop derivatives:
\begin{equation}\label{eq:C10}
\frac{\partial N_\alpha}{\partial\lambda_\beta}=\sum_I\left(A_{\lambda_\beta}\right)_I 2c_IN^\alpha_I.
\end{equation}

\end{document}